\documentclass[multicol,showpacs,aps,draft]{revtex4}

\usepackage{graphicx}% Include figure files
\usepackage{dcolumn}% Align table columns on decimal point
\usepackage{bm}% bold math
\usepackage{feynmf}
\usepackage{psfig}

\newcommand{\la}{\left\langle}
\newcommand{\ra}{\right\rangle}
\newcommand{\be}{\begin{equation}}
\newcommand{\ee}{\end{equation}}
\newcommand{\bea}{\begin{eqnarray}}
\newcommand{\eea}{\end{eqnarray}}
\newcommand{\ba}{\begin{array}}
\newcommand{\ea}{\end{array}}

\begin{document}
\title{Field theoretic calculation of scalar turbulence}
\author{Mahendra\ K.\ Verma  \thanks{email: mkv@iitk.ac.in}}
\affiliation{Department of Physics, Indian Institute of Technology,
Kanpur  --  208016, INDIA}
\date{24 August 2001}

\begin{abstract}

The cascade rate of passive scalar and Bachelor's constant in scalar
turbulence are calculated using the flux formula.  This calculation is
done to first order in perturbation series.  Batchelor's constant in
three dimension is found to be approximately 1.25.  In higher dimension,
the constant increases as $d^{1/3}$.
\end{abstract}

\vspace{1.5cm}

% isotropic turbulence; turbulent diffusion; field theory
\maketitle

\section{Introduction}

	Perturbative field-theoretic techniques have been very
useful in turbulence research.  One of the celebrated field-theoretic
method, renormalization groups (RG), has been applied to fluid
turbulence \cite{FNS,YakhOrsz,McCo:book}, scalar turbulence
\cite{YakhOrsz,Zhou:scalar}, MHD turbulence \cite{MKV:MHD_PRE} etc.  In
RG analysis, one can calculate the renormalized parameters at large
length scales.  In addition to RG, one can also apply the
field-theoretic techniques to calculate turbulent cascade rates
\cite{Lesl:book,MKV:MHD_PRE}.  In this paper we will calculate the
cascade rates of passive scalar using perturbative technique.  From
this calculation we can also calculate Batchelor's constant, which is
very important for large-eddy simulations.

The study of passive scalar is one of the important areas in
turbulence research.  It finds application in evolution of temperature
field, pollution diffusion, etc.  The phenomenology of passive scalar
is well developed \cite{Lesl:book}, and their predictions are in
agreement with the experimental results.  According to the
phenomenology, the energy spectrum of both velocity field ${\bf u}$
and scalar field $\psi$ in the inertial-convective range are
proportional to $k^{-5/3}$.  Note that in the inertial-convective
range both the nonlinear terms ${\bf u \cdot \nabla u}$ and ${\bf u
\cdot \nabla \psi}$ dominate the viscous term.  However, there exist
two other ranges depending on the value of Prandtl number (the ratio
of viscosity and diffusivity).  In this paper we will only focus on
inertial-convective range.

Regarding the calculation of renormalized viscosity and diffusivity
for passive scalar admixture, Yakhot and Orszag \cite{YakhOrsz}
adopted $\epsilon$-expansion, while Zhou and Vahala
\cite{Zhou:scalar}, and Lin et al.'s~\cite{Lin} procedure is recursive
based on the original idea of McComb and his group (\cite{McCo:book}
and reference therein).  Adzhemyan et al.~\cite{Adzh} used De
Dominicis and Martin's \cite{DeDo} procedure for fluid turbulence to
passive scalars and computed the renormalized parameters.  Earlier,
Wyld \cite{Wyld} had given a perturbative expansion of Navier-Stokes
equation.  Canuto and Dubovikov \cite{Canu1}, and Canuto et
al.~\cite{Canu3} started with Wyld's formalism and computed the
renormalized diffusivity for passive scalar; they also computed
Batchelor's constant. 

Turbulence cascade rates or fluxes play important role in turbulence
calculations. It is a measure of transfer of a certain quantity from
inside of a wavenumber sphere to the outside wavenumber sphere.  In
fluid turbulence, Kraichnan \cite{Krai:59} applied direct interaction
approximation and calculated the flux.  Later, the cascade rates have
been calculated by many researchers using various techniques,
e.g. Eddy-Damped-Quasi-Normal-Markovian (EDQNM) closure scheme, RG,
etc.  Here we are interested in the cascade rate of passive scalar.
This cascade rate quantifies how scalar fluctuations at large length-scales
diffuse to small length-scales.

In this paper we apply perturbative techniques to calculate the
cascade rate of passive scalar.  In our scheme the cascade rate of
passive scalar is calculated using the flux formula and the
renormalized parameters.  In section 2 we recapitulate the earlier RG
calculation \cite{Zhou:scalar} and extend their results to higher
dimensions.  In the subsequent section we apply the perturbative
technique and calculate the cascade rate in the inertial-convective
range to first order. The final expression involves energy spectrum
for which we substitute $k^{-5/3}$ obtained from the phenomenology.
From this procedure we also calculate Batchelor's constant.  We have
extended our calculations to higher space dimensions, because 
higher-dimensional field theory usually provide important insights into the
nature of nonlinear interactions \cite{Nelk}.

The outline of the paper is as follows: in section 2 we provide the
definitions and recapitulation of the renormalization procedure for
passive scalar. In section 3, we carry out the calculation of flux of
passive scalar and Batchelor's constant.  Section 4 contains
conclusions.

\section{renormalization of viscosity and diffusivity 
	revisited}

Earlier calculations of renormalization in scalar turbulence have been
carried out by Yakhot and Orszag \cite{YakhOrsz}, Zhou and Vahala
\cite{Zhou:scalar}, and Lin et al.~\cite{Lin}.  In this section we
recapitulate very briefly Zhou and Vahala's calculation for passive
scalar and extend their results to higher dimensions.  Zhou and
Vahala's calculation is based on recursive scheme proposed by McComb
and his coworker (\cite{McCo:book} and references therein).  The
equations for the velocity {\bf u} and passive scalar $\psi$ fields in
Fourier space are
\bea 
\left( -i\omega + \nu k^2 \right) u_i (\hat{k})  & = & 
 -\frac{i}{2} P_{ijm}({\bf k}) \int d\hat{p}  u_j (\hat{p}) 
u_m (\hat{k}-\hat{p}) 
\label{eqn:NSk} \\ 
\left( -i\omega + \kappa   k^2 \right) \psi (\hat{k}) & = & 
 -i k_j \int d\hat{p}  u_j (\hat{p}) \psi (\hat{k}-\hat{p}) 
\label{eqn:scalark}
\eea
with 
\bea 
P_{ijm}({\bf k})   & =  & k_j P_{im}({\bf k}) - k_m P_{ij}({\bf k}); \\
P_{im}({\bf k})  & =  & \delta_{im}-\frac{k_i k_m}{k^2}; \\
\hat{k} & = & ({\bf k},\omega); \\ 
d \hat{p} & = & d {\bf p} d \omega/(2 \pi)^{d+1}. 
\eea 
Here  $\nu$   and    $\kappa$ are   the    viscosity  and  diffusivity
respectively,  $p$  is  the  fluid pressure,  and   $d$ is  the  space
dimension.  We have assumed that the flow is incompressible, i.e. $k_i
u_i({\bf k}) = 0$.

	In the recursive RG procedure the wavenumber range $(k_N,k_0)$
is divided logarithmically into N shells.  The effective parameters are
obtained by eliminating the high wavenumber shells
iteratively.  We denote the higher wavenumber shells by $k^>$ and the
remaining wavenumber region by $k^<$.  In this procedure the field
variables $u^>_i(\hat{k})$ and $\psi^>(\hat{k})$ are assumed to be
gaussian with zero mean, and
\bea
\left\langle u_i^> (\hat{p}) u_j^> (\hat{q})\right\rangle  & = &
P_{ij}({\bf p)} C^{u} (\hat{p}) \delta(\hat{p}+\hat{q}) 
\label{eqn:Cu}\\
\left\langle \psi^> (\hat{p}) \psi^> (\hat{q})\right\rangle  &= &
 C^{\psi} (\hat{p}) \delta(\hat{p}+\hat{q}) \label{eqn:Cpsi}
\eea
where $C^{u}(\hat{p})$ and $C^{\psi}(\hat{p})$ are velocity
and scalar correlation functions respectively.

If we denote $\nu_{(n)}$ and $\kappa_{(n)}$ as viscosity and
diffusivity respectively after the elimination of $n$ shells, then the
elimination of the next shell yields the following equations to the first
order in perturbation:

\bea 
\left( -i\omega + \nu_{(n)} k^2 + 
	\delta \nu_{(n)} k^2 \right) u_i^<(\hat{k}) & = & 
	-\frac{i}{2} P_{ijm}({\bf k}) \int d\hat{p}
	u_j^< (\hat{p}) u_m^< (\hat{k}-\hat{p})  \\
\left( -i\omega + \kappa_{(n)} k^2 
	+ \delta \kappa_{(n)} k^2 \right) \psi^<(\hat{k}) & = &  
	-i k_j \int d\hat{p}  u_j^< (\hat{p}) \psi^< (\hat{k}-\hat{p}) 
\eea 
where
\input{rgfeyn.diag}

In the above Feynmann diagrams, the solid, wiggly (photon), and curly
(gluon) lines represent correlation function $\la u_i u_j \ra$, and
Green functions $G^u$, $G^{\psi}$ respectively.  The filled circle
represents $(-i/2) P_{ijm}$ vertex, while the empty circle
represents $-i k_j$ vertex.  The RG procedure adopted here is the same
as that of Zhou and Vahala \cite{Zhou:scalar}.  Some of the notation
used here is close to the that of MHD turbulence calculation of Verma
\cite{MKV:MHD_PRE,MKV:MHDRG}.  

The frequency dependence of the correlation function are taken as:
$C^{u}(k,\omega)=2 C^{u}(k) \Re(G^{u}(k,\omega))$ and
$C^{\psi}(k,\omega)=2 C^{\psi}(k) \Re(G^{\psi}(k,\omega))$.  With this
assumption, the expressions corresponding to the above Feynmann
diagrams will be 
\bea 
\delta \nu_{(n)}(k) & = & \frac{1}{(d-1)k^2} 
			  \int^{\Delta}_{\bf p+q=k}
				 \frac{d {\bf p}}{(2 \pi)^d}
 			  S_1(k,p,q) \frac{C^{u}(q)}
			  {\nu_{(n)}(p) p^2+\nu_{(n)}(q) q^2}
                         \label{eqn:nu} \\ 
\delta \kappa_{(n)}(k) & = & \frac{1}{k^2}
			      \int^{\Delta}_{\bf p+q=k}
				 \frac{d {\bf p}}{(2 \pi)^d}
			    S_2(k,p,q) \frac{C^{u}(q)}
			      {\kappa_{(n)}(p) p^2+\nu_{(n)}(q) q^2}
                             \label{eqn:kappa} 
\eea
with
\bea
S_1(k,p,q) & = & kp \left( (d-3)z+2 z^3+(d-1)xy \right) \\ 
S_2(k,p,q) & = & k p \left( z+x y \right) 
\eea
The quantities $x,y,$ and $z$ are defined by
\be
x= - \frac{{\bf p \cdot q}}{pq}; \hspace{1cm} y=\frac{{\bf q \cdot k}}{qk}; 
\hspace{1cm}  z=\frac{{\bf p \cdot k}}{pk}.
\ee
The effective viscosity and diffusivity after the elimination
of $(n+1)$ shell are
\bea
(\nu,\kappa)_{(n+1)} (k) & = & (\nu,\kappa)_{(n)} (k) +
			\delta (\nu,\kappa)_{(n)} (k) 
			\label{eqn:nukappa_n} 
\eea

The spectrum $C^{u}(k)$ can be written in terms of 
one-dimensional energy spectrum $E^{u}(k)$ as
\be
C^{u}(k) = \frac{2 (2 \pi)^d}{S_d (d-1)} k^{-(d-1)} E^{u}(k)
	\label{eqn:Cu_k} 
\ee
where $S_d$ is the surface area of $d$ dimensional spheres.  It is known
that $E^u(k)$ follows Kolmogorov's spectrum, i.e.,
\be
	E^u(k)  =  K^u (\Pi^u)^{2/3} k^{-5/3}
	\label{eqn:Eu_k} 
\ee
where $\Pi$ is the kinetic-energy flux, and $K^u$ is 
Kolmogorov's constant for fluid turbulence.
Using the dimensional arguments we find that $\nu_{(n)}$ and
$\kappa_{(n)}$ have the following forms:
\bea
(\nu,\kappa)_{(n)} (k_n k') & = & (K^u)^{1/2} (\Pi^u)^{1/3} k_n^{-4/3} 
			(\nu,\kappa)_{(n)}^* (k') 
\eea
with $k=k_{n+1}k' (k' < 1)$.  The large-$n$ limit of the 
$\nu_{(n)}^* (k')$ and $\kappa_{(n)}^* (k')$ are expected to
be universal functions in the RG sense.

We solve for $\nu_{(n)}^*(k')$ and $\kappa_{(n)}^*(k')$ iteratively
using Eqs.~(\ref{eqn:nu}, \ref{eqn:kappa}, \ref{eqn:nukappa_n}).  We
take $h=0.7$, and start with constant $\nu_{(0)}^*$ and
$\kappa_{(0)}^*$. We iterate the process till $\nu^*_{(n+1)}(k')
\approx \nu^*_{(n)}(k')$ and $\kappa^*_{(n+1)}(k') \approx
\kappa^*_{(n)}(k')$, that is, till they converge.  We find that the
iteration process converges; the limiting value $\nu^*$ and
$\kappa^*$ are shown in Table \ref{tab:scalar}.

We can draw many interesting conclusions from the above results.
Since the scalar does not appear in the equation for $u$, $\nu^*$
computed here is the same as that obtained for fluid turbulence.  In
Table \ref{tab:scalar} we have listed the renormalized diffusivity
$\kappa^*$ and the turbulent Prandtl number $Pr_{turb}$.  For $d=3$,
$\kappa^*=0.85$ and $Pr_{turb} = \nu^*/\kappa^*=0.42$.  The above
quantities vary a bit with the variation of $h$, but they are roughly
in the same range.  The error in our estimate of the parameters is of
the order of 0.1.  Our results are in the same range as those obtained
by Zhou and Vahala \cite{Zhou:scalar}.

We have also carried out the above analysis for higher space
dimensions.  The calculated $\kappa^*$ and $Pr_{turb}$ are listed in
Table \ref{tab:scalar}.  For large $d$, $\nu^* \approx \kappa^*
\propto d^{-1/2}$.  The $d$ dependence is in the agreement with the
finding of Fournier and Frisch for fluid turbulence \cite{FourFris}.
The above result also implies that $Pr_{turb} \approx 1$ for large
$d$.

In two-dimensions the scalars are not constrained to double
energy-enstrophy conservation like velocity field.  The RG analysis
for two-dimensional scalar turbulence is beyond the scope of this
paper.

\section{Calculation of cascade rates}
	
In this section we compute cascade rates of $u$ and $\psi$, and
Bachelor's constant.  To this end we use the flux formulas and the
renormalized parameters computed in the previous section.  The time
evolution of correlation functions $C^u$ and $C^{\psi}$ (defined by
Eqs. [\ref{eqn:Cu}, \ref{eqn:Cpsi}]) are given by
\cite{Lesl:book,Stan:book,MKV:MHDflux,Dar:flux}
\bea
\left(\frac{\partial}{\partial t}  + 2 \nu k^2 \right) 
			C^{u}({\bf k},t,t) 
& = &	\frac{1}{(d-1)(2 \pi)^d \delta({\bf k+k'})} 
	\int_{\bf k'+p+q=0} \frac{d {\bf p}}{(2 \pi)^d} 
	       [S^{uu}({\bf k'|p|q})+S^{uu}({\bf k'|q|p})]  
			\label{eqn:Cu_t} \\
\left(\frac{\partial}{\partial t}  + 2 \kappa k^2 \right) 
			C^{\psi}({\bf k},t,t) 
& = &	\frac{1}{(2 \pi)^d \delta({\bf k+k'})} 
	\int_{\bf k'+p+q=0} \frac{d {\bf p}}{(2 \pi)^d} 
	       [S^{\psi \psi}({\bf k'|p|q})+S^{\psi \psi}({\bf k'|q|p})]
			\label{eqn:Cpsi_t} 
\eea
where
\bea
S^{uu}({\bf k'|p|q}) &  = & -\Im \left({\bf \left[k'.u(q)\right]} %
{\bf \left[u(k').u(p)\right]} \right)  \label{eq:Sukup_def} \\
S^{\psi \psi}({\bf k'|p|q}) &  = & -\Im \left({\bf \left[k'.u(q)\right]}
 \left[\psi(k') \psi(p) \right] \right)  \label{eq:Spsikpsip_def} 
\eea
Here $\Im$ stands for the imaginary part of the argument.  Note that
Eqs.~(\ref{eqn:Cu_t}, \ref{eqn:Cpsi_t}) have been discussed in the
earlier literature, e.g., Lesieur \cite{Lesl:book} and
Stani\u{s}i\'{c} \cite{Stan:book}.  However, reinterpretation of the
terms $S({\bf k|p|q})$ by Dar et al.~\cite{Dar:flux} as energy transfer from
mode {\bf p} (the second argument of $S$) to {\bf k} (the first
argument of $S$) with mode {\bf q} (the third argument of $S$) as a
mediator makes the formalism more transparent and simple.  Also, some
quantities which were impossible to calculate in earlier formalism
could be computed now \cite{Dar:flux}.  This interpretation of Dar et
al. is consistent with the earlier formalism.  

The energy fluxes $\Pi^u$ and $\Pi^{\psi}$ from a wavenumber sphere of
radius $k_0$ is \cite{Dar:flux}
\bea
\Pi^{u}(k_0) & = & \int_{k'>k_0} \frac{d {\bf k'}}{(2 \pi)^d} 
		       \int_{p<k_0} \frac{d {\bf p}}{(2 \pi)^d}  
			\la S^{uu}({\bf k'|p|q}) \ra 
			\label{eqn:u_flux}	\\
\Pi^{\psi}(k_0) & = & \int_{k'>k_0} \frac{d {\bf k'}}{(2 \pi)^d} 
		       \int_{p<k_0} \frac{d {\bf p}}{(2 \pi)^d}  
			\la S^{\psi \psi}({\bf k'|p|q}) \ra 
			\label{eqn:psi_flux}	
\eea
Note that there is no cross-transfer between $u$ and $\psi$ energy.
It is also important to note that both $C^u$ and $C^{\psi}$ are
conserved in every triad interaction, i.e.,
\bea
S^{uu}({\bf k'|p|q}) + S^{uu}({\bf k'|q|p}) + 
S^{uu}({\bf p|k'|q}) + S^{uu}({\bf p|q|k'}) + 
S^{uu}({\bf q|k'|p}) + S^{uu}({\bf q|p|k'}) & = & 0  \\
S^{\psi \psi}({\bf k'|p|q}) + S^{\psi \psi}({\bf k'|q|p}) + 
S^{\psi \psi}({\bf p|k'|q}) + S^{\psi \psi}({\bf p|q|k'}) + 
S^{\psi \psi}({\bf q|k'|p}) + S^{\psi \psi}({\bf q|p|k'}) & = & 0
\eea
These are the statements of  ``detailed conservation of energy''
in triad interaction  (when $\nu=\kappa=0$) \cite{Lesl:book}.

The energy fluxes can be calculated using Eqs. (\ref{eqn:u_flux},
\ref{eqn:psi_flux}) by taking ensemble averages of $S^{uu}$ and
$S^{\psi \psi}$.  It is easy to check that $\la S^{uu} \ra = \la
S^{\psi \psi} \ra =0$ to the zeroth order, but are nonzero to the
first order.  The field-theoretic calculation performed here is very
similar to Verma's MHD flux calculation \cite{MKV:MHDflux}.  Please
refer to Verma's paper \cite{MKV:MHDflux} for further details.  The
Feynmann diagrams for the first order of $\la S \ra$ are
\vspace{0.25cm}

\input{energyfeyn.diag}

\vspace{0.25cm}

In the above Feynmann diagrams, the solid, dashed, wiggly (photon),
and curly (gluon) lines represent $\la u_i u_j \ra$, $\la \psi \psi
\ra$,  $G^u$, and $G^{\psi}$ respectively.  In all the diagrams,
the left vertex denotes $k_i$, while the filled circle and the empty
circles of right vertex represent $(-i/2) P_{ijm}$ and $-i k_j$
respectively.  Algebraically,
\bea 
\la S^{uu}(k|p|q)\ra & = &\int_{-\infty}^t  dt'  [
                T_{1}(k,p,q) G^{u}(k,t-t') C^{u}(p,t,t') C^{u}(q,t,t') 
			\nonumber \\
& &\hspace{1cm} + T_{2}(k,p,q) G^{u}(p,t-t') C^{u}(k,t,t') C^{u}(q,t,t') 
			\nonumber \\
& &\hspace{1cm} + T_{3}(k,p,q) G^{u}(q,t-t') C^{u}(k,t,t') C^{u}(p,t,t') ]
			\label{eqn:Suu}	\\
\la S^{\psi \psi}(k|p|q)\ra & = &\int_{-\infty}^t  dt'  [
                T_{4}(k,p,q) G^{\psi}(k,t-t') C^{\psi}(p,t,t') 
			C^{u}(q,t,t') 
			\nonumber \\
& &\hspace{1cm} + T_{5}(k,p,q) G^{\psi}(p,t-t') C^{\psi}(k,t,t') 
			C^{u}(q,t,t') ]
			\label{eqn:Spsipsi}
\eea
where $T_i(k,p,q)$'s are given by
\bea
T_1(k,p,q) & = &  -kp \left( (d-3)z + (d-2)xy +2 z^3+ 2 x y z^2
			  + x^2 z \right) \\
T_2(k,p,q) & = &   kp \left( (d-3)z + (d-2)xy +2 z^3+ 2 x y z^2
			  + y^2 z \right) \\
T_3(k,p,q) & = &   kq \left(x z - 2 x y^2 z - y z^2 \right)  \\
T_4(k,p,q) & = &   k^2 \left(1 - y^2 \right)  \\
T_5(k,p,q) & = &   -k p \left( z + xy \right)
\eea

We assume the relaxation time for $C^u(k)$ and $C^{\psi}(k)$
to be $(\nu(k) k^2)^{-1}$ and $(\kappa(k) k^2)^{-1}$ respectively,
i.e.,
\bea
C^{u}(k,t,t') & = & \exp \left(- \nu(k) k^2 (t-t') \right) 
			C^{u}(k,t,t) \\ 
C^{\psi}(k,t,t') & = & \exp \left(- \kappa(k) k^2 (t-t') \right) 
			C^{\psi}(k,t,t) 
\eea
With this assumption, 
Eqs.~(\ref{eqn:Suu}, \ref{eqn:Spsipsi}) reduce to
\bea
\Pi^{u}(k_0) & = & \int_{k>k_0} \frac{d {\bf k}}{(2 \pi)^d} \int_{p<k_0} 
                         \frac{d {\bf p}}{(2 \pi)^d}  
   	                 \frac{1}{\nu(k) k^{2}+ 
				\nu(p) p^{2}+\nu(q) q^{2}}
			\times \nonumber \\
& &  [ T_{1}(k,p,q) C^{u}(p) C^{u}(q)
    +T_{2}(k,p,q) C^{u}(k) C^{u}(q)
     +T_{3}(k,p,q) C^{u}(k) C^{u}(p) ] 
    \label{eqn:Pi_u}  \\
\Pi^{\psi}(k_0) & = & \int_{k>k_0} \frac{d {\bf k}}{(2 \pi)^d} \int_{p<k_0} 
                         \frac{d {\bf p}}{(2 \pi)^d}  
   	                 \frac{1}{\kappa(k) k^{2}+ 
				\kappa(p) p^{2}+\nu(q) q^{2}}
			\times \nonumber \\
& &  [ T_{4}(k,p,q) C^{\psi}(p) C^{u}(q)
    +T_{5}(k,p,q) C^{\psi}(k) C^{u}(q) ]
    \label{eqn:Pi_psi}  
\eea
For $C^u(k)$ we substitute Eqs.~(\ref{eqn:Cu_k}, \ref{eqn:Eu_k}), 
while for $C^{\psi}$ we substitute \cite{Lesl:book}
\bea
C^{\psi}(k) & = & \frac{2 (2 \pi)^d}{S_d} k^{-(d-1)} E^{u}(k), \\
	\label{eqn:Cpsi_k}  
E^{\psi}(k) & = & K^{\psi} \Pi^{\psi} (\Pi^u)^{-1/3} k^{-5/3}
	\label{eqn:Epsi_k} 
\eea
where $K^{\psi}$ is called the Batchelor's constant.  The renormalized
viscosity and diffusivity in the inertial range are
\bea
\nu(k) & = & (K^u)^{1/2} (\Pi^u)^{1/3} k^{-4/3} \nu^*    
	 \label{eqn:nuk} \\
\kappa(k) & = & (K^u)^{1/2} (\Pi^u)^{1/3} k^{-4/3} \kappa^*  .
	\label{eqn:kappak}
\eea
The substitution of the above quantities, and the change of variables
\be
k=\frac{k_0}{u}; p=\frac{k_0}{u} v; q=\frac{k_0}{u} w
\ee
yield the following nondimensional version of the flux equations
\cite{FourFris}:
\bea
1 & = &  (K^u)^{3/2} \left[ \frac{4 S_{d-1}}{(d-1)^2 S_d}
               \int_0^1 dv \ln{(1/v)} \int_{1-v}^{1+v} dw 
		(vw)^{d-2} (\sin \alpha)^{d-3}
		F^u(v,w) \right]
\label{eqn:Piu} \\
1 & = & K^{\psi} (K^u)^{1/2} \left[ \frac{4 S_{d-1}}{(d-1) S_d}
               \int_0^1 dv \ln{(1/v)} \int_{1-v}^{1+v} dw 
		(vw)^{d-2} (\sin \alpha)^{d-3}
		F^{\psi}(v,w) \right]
\label{eqn:Pipsi}
\eea
where $\alpha$ is angle between vectors ${\bf p}$ and ${\bf q}$,
and the integrals $F^{u,\psi}(v,w)$ are
\bea
F^{u} & = &  \frac{1}{\nu^*(1+v^{2/3}+w^{2/3})} 
		[t_{1}(v,w) (v w)^{-d-\frac{2}{3}}
		+t_{2}(v,w) w^{-d-\frac{2}{3}} 
	        +t_{3}(v,w) v^{-d-\frac{2}{3}} ] \\
F^{\psi} & = &  \frac{1}{\kappa^* (1+v^{2/3})+\nu^* w^{2/3}} 
		[t_{4}(v,w) (v w)^{-d-\frac{2}{3}}
		+t_{5}(v,w) w^{-d-\frac{2}{3}} ]
\eea
Here $t_i(v,w) = T_i(k,kv,kw)/k^2$.  

The terms in the square brackets of Eq.~(\ref{eqn:Piu}, \ref{eqn:Pipsi})
(denoted by $I^{u,\psi}$) involve integrals.  We compute them using
gaussian quadrature.  The integrals converge for all dimensions $d \ge
2$.  Once the integrals are known, Kolmogorov's and Batchelor's
constants ($K^u$ and $K^{\psi}$ respectively) can be computed.  The
computed values are given in Table 1.

In our calculation Batchelor's constant $K^{\psi}$ in three dimension
is 1.25.  Due to uncertainties in the value of $\nu^*$ and $\kappa^*$,
the error in the constant could be of the order of 0.1.  Earlier,
Kraichnan had estimated the constant to be 0.2. Yakhot and Orszag
\cite{YakhOrsz} obtained $K^{\psi}=1.16$ by their $\epsilon$-based
renormalization group analysis.  Canuto and Dubovikov \cite{Canu1} and
Canuto et al.~\cite{Canu3} estimated $K^{\psi} = (5/3)*0.72 = 1.2$
using their RG calculation.  Lin et al.~\cite{Lin} find the constant
to be close to 0.3.  Our result is in very good agreement with the
theoretical predictions of Yakhot and Orszag \cite{YakhOrsz} and
Canuto et al.~\cite{Canu3}, as well as to the experimental values
($\approx 1.2-1.4$, see Monin and Yaglom \cite{MoniYagl2:book}).

It is also interesting to note that both $K^{u,\psi}$ are proportional
to $d^{1/3}$, consistent with the predictions of Fournier and Frisch
\cite{FourFris} for fluid turbulence.  This result implies that the
cascade rated $\Pi^{u,\psi}$ will decrease with dimensions as
$d^{-1/2}$.

\section{Conclusions}
	
In this paper we employed field-theoretic techniques to calculate the
cascade rates of scalar turbulence.  Our calculation is to first
order. From this formalism we also calculate Batchelor's constant.  In
three dimensions, we find Batchelor's constant to be 1.25, which is in
very good agreement with the theoretical predictions of Yakhot and
Orszag \cite{YakhOrsz} and Canuto et al.~\cite{Canu3}, and the
experimental values.  In higher space dimensions the constant varies
as $d^{1/3}$.

Our calculation of cascade rate requires the renormalized viscosity
and diffusivity.  We have extended the RG calculations of Zhou and
Vahala \cite{Zhou:scalar} for higher dimensions.  Our calculations
show that for higher dimensions, the renormalized viscosity and
diffusivity vary with dimensions as $d^{-1/2}$, and the turbulent
Prandtl number approaches unity.

%\bibliographystyle{prsty}
%\bibliography{/home/mkv/mypapers/bib/abbrev,/home/mkv/mypapers/bib/surf,/home/mkv/mypapers/bib/burg,/home/mkv/mypapers/bib/mhd,/home/mkv/mypapers/bib/fluid,/home/mkv/mypapers/bib/misc,/home/mkv/mypapers/bib/mkv}

\begin{table}
\caption{The computed values of renormalized viscosity $\nu^*$,
diffusivity $\kappa^*$, turbulent Prandtl number $Pr_{turb}$,
Kolmogorov's constant $K^u$ and Batchelor's constants $K^{\psi}$ for
various space dimensions $d$.}

\label{tab:scalar}
\begin{ruledtabular} 
\begin{tabular}{lccccr} 
$d$    & $\nu^*$ & $\kappa^*$ & $Pr_{turb}$ & $K^u$  & $K^{\psi}$  \\  \hline
3      &   0.36  &  0.85      & 0.42        & 1.53   & 1.25  \\
4      &   0.42  &  0.69      & 0.61        & 1.60   & 1.39  \\
7      &   0.38  &  0.48      & 0.80        & 1.76   & 1.65  \\
10     &   0.34  &  0.39      & 0.87        & 1.94   & 1.83  \\
25     &   0.22  &  0.24      & 0.94        & 2.43   & 2.44  \\
50     &   0.16  &  0.16      & 1.0           & 3.1    & 3.0  \\
100    &   0.093 & 0.095      & 0.98        & 3.4    & 3.4  \\
\end{tabular}
\end{ruledtabular}
\end{table}

\end{document}